# Electrical Transport Across an Individual Magnetic Domain Wall in (Ga,Mn)As Microdevices

Hongxing Tang and Michael L. Roukes

*Condensed Matter Physics 114-36, California Institute of Technology, Pasadena, CA 91125, USA*

Recent studies demonstrate that an individual magnetic domain wall (DW) can be trapped and reproducibly positioned within multiterminal (Ga,Mn)As microdevices. The electrical resistance obtained from such measurements is found to be measurably altered by the presence of this single entity. To elucidate these observations we develop a simple model for the electrical potential distribution along a multiterminal device in the presence of a single DW. This is employed to calculate the effect of a single DW upon the longitudinal and transverse resistance. The model provides very good agreement with experimental observations, and serves to highlight important deviations from simple theory. We show that measurements of transverse resistance along the channel permits establishing the position and the shape of the DW contained within it. An experimental scheme is developed that enables unambiguous extraction of the *intrinsic* DW resistivity. This permits the intrinsic contribution to be differentiated from resistivities originating from the bulk and from magnetic anisotropy – effects that are generally manifested as large backgrounds in the experiments.

The electrical resistance arising from a single magnetic domain wall (DW) has been of interest for many years.[1] Recent developments in ferromagnetic semiconductors have renewed interest in this area. The unique spin configuration across a DW is similar to the spin alignment underlying the giant magnetoresistance effect, and it is also central to recent concepts for domain wall spin transistors.[2,3]

In homogeneous ferromagnets electrical transport is affected both by classical magnetoresistance phenomena such as Lorenz force induced magnetoresistance (LMR), and phenomena arising from the presence of the spontaneous magnetization such as the Anomalous Hall resistance (AHR) and the anisotropic magnetoresistance (AMR). A number of recent theoretical and experimental investigations have focused upon the nature of the resistance arising from a domain wall itself. Distinguishing between these different contributions to the resistance, which are all simultaneously present in real experiments, poses a significant challenge.

There are two principal issues that complicate differentiation between domain wall effects and bulk phenomena. First, ideal observations would involve a few or, ideally, just a single domain wall to unambiguously isolate its effects from those of others. Second, one also wishes to separate simple "classical" phenomena – those which can arise solely from the resistivity discontinuity at a domain wall – from smaller, more subtle *magnetic* scattering phenomena in that same locale. Hereafter, we term the latter contributions as the "intrinsic" domain wall resistivity.

In order to obtain simple domain patterns and to avoid extrinsic magnetoresistance contribution, recent experiments have concentrated on studying domain walls in narrow, submicron-width ferromagnetic metal wires or nanoconstrictions. However, at nanometer length scales, it has proven difficult to extract the intrinsic magnetoresistance of domain walls because both the domain wall structure and current flow can be significantly altered by the complex geometry and magnetic anisotropy.[4] These complications have precluded high precision measurements of the intrinsic domain wall resistance (DWR). Even for 3D metals, in which many experimental studies have been performed, a clear understanding of the observed phenomena remains elusive; both negative and positive intrinsic DWRs have been reported.[5,6,7,8,9,10,11] The theoretical results are equally ambiguous. Several semiclassical scattering mechanisms predict *positive* DWR: these include (a) reflection of carriers by the domain wall,[12] (b) zigzag current redistribution inside the wall due to Hall effect,[13] and (c) spin-dependent scattering analogous to the GMR effect in magnetic multilayers.[14,15] But the possibility of *negative* domain wall resistance has also emerged, and explained in the context of electronic coherence in ferromagnetic metals. It has been shown that domain wall scattering can lead to suppression of the dephasing in a weakly localized system, which in turn can reduce the resistivity of domain walls. This source of negative domain wall resistance originates from quantum mechanical corrections.[16]

Epitaxial films of the ferromagnetic semiconductor (Ga,Mn)As demonstrate extremely simple domain structures even at macroscopic length scales.[17,18] Multiterminal devices patterned from these epilayers have recently enabled direct electrical measurements upon individual domain walls. Here we develop an analytical model of electrical transport in such devices, to evaluate the experimental manifestations that are expected when a single domain wall is present. We show that the current distributions become significantly modified in the locale of the domain wall. The evolution of transverse and longitudinal resistivities as an individual domain wall propagates through the device is calculated. We find extremely good agreement between theoretical predictions and experimental data, with subtle differences emerging that highlight the role of the intrinsic domain wall resistivity. In fact, this simple model establishes an unambiguous experimental protocol for the extraction of the intrinsic domain wall resistance from larger bulk magnetoresistance effects.

This paper is organized as follows. Section I presents our simple model based upon the assumption of a local resistivity tensor, which describes current flow in the presence of a electrical resistivity discontinuity associated with a domain wall. Section II describes the "eddy-like" currents that result from such a static domain wall positioned between probes within the sample. Section III describes the transverse resistance that is generated by a domain wall within the sample. Section IV presents the resulting longitudinal resistance, and a protocol for differentiating the contribution arising from the "intrinsic" domain wall resistance from bulk effects. Section V summarizes our most important conclusions.

## I. Modeling of current flow in the presence of a magnetic domain wall

Figure 1a shows a typical multiprobe device such as employed in recent experiments.[18,19] In these experiments, four-probe transverse resistances are measured using pairs of voltage probes located across the device channel, and four-probe longitudinal resistances are measured using probe pairs located on the top and bottom sides of the channel. Referring to Fig. 1a, the former are represented as $R_{xy}^L \equiv R_{14,26}$ and $R_{xy}^R \equiv R_{14,35}$, and the latter as $R_{xx}^U \equiv R_{14,23}$ and $R_{xx}^D \equiv R_{14,65}$, respectively. Here the superscripts refer to left and right for the transverse resistances (subscripts $xy$), and up and down for the longitudinal resistances (subscripts $xx$). These resistances have been succinctly defined above using conventional four-probe notation; where $R_{ij,kl}$ corresponding to a sensing current imposed from terminal $i$ to $j$, which results in an induced potential from $k$ to $l$.

To capture the essential features arising from the presence of a single domain wall within the device we model it simply; we assume the domain wall is straight, oriented with arbitrary angle with respect to the channel boundaries, and translates along the device channel without changing its shape. This configuration depicted in Figure 1b: we define the $x$ axis along the device channel and the $y$ axis perpendicular to it. The domain wall is located at position $x_0$ and has slope $k$. We assume that the easy axis is oriented along angle $\varphi_1$ prior to reversal, while after reversal is along $\varphi_2$. For the case of purely cubic anisotropy, $\varphi_1 + \varphi_2 = 90^o$. If the device channel is aligned precisely along the hard axis, $\varphi_1 = -\varphi_2$. In the most general case, however, this is not satisfied, therefore these two magnetization orientation across the domain wall are more generally be represented as,

$$\varphi(x) = \phi_0 \operatorname{sgn}(x - x_0 - y/k) + \delta \qquad (1)$$

This is depicted in Fig. 1b. For purely cubic magnetic anisotropy $\phi_0 = 45^o$, however in our recent experiments with (Ga,Mn)As epilayers, a weak in-plane uniaxial anisotropy is found that is superimposed along [110] direction. This results in a value $\phi_0 = 37°$.[18] The misalignment from [110] orientation is denoted by angle $\delta$.

In (Ga,Mn)As, the resistivity in the direction along the magnetization, $\rho_{//}$, is smaller than in the direction perpendicular to it, $\rho_{\perp}$. The resistivity tensor in our coordinate system can be calculated directly from the resistivity in the diagonal frame,

$$
\begin{aligned}
\rho_{\varphi} &= \begin{pmatrix} \cos\varphi & -\sin\varphi \\ \sin\varphi & \cos\varphi \end{pmatrix} \begin{pmatrix} \rho_{//} & 0 \\ 0 & \rho_{\perp} \end{pmatrix} \begin{pmatrix} \cos\varphi & \sin\varphi \\ -\sin\varphi & \cos\varphi \end{pmatrix} \\
&= \begin{pmatrix} \bar{\rho} + \frac{1}{2}\Delta\rho\cos 2\varphi & \frac{1}{2}\Delta\rho\sin 2\varphi \\ \frac{1}{2}\Delta\rho\sin 2\varphi & \bar{\rho} - \frac{1}{2}\Delta\rho\cos 2\varphi \end{pmatrix}
\end{aligned}
\tag{2}
$$

in which $\bar{\rho} = (\rho_{//} + \rho_{\perp})/2$. In the optimum situation, i.e., precise alignment is achieved and the magnetic anisotropy is strictly cubic, the resistivity tensor can be simplified,

$$
\rho_{I,II} = \bar{\rho} \begin{pmatrix} 1 & \pm\frac{1}{2}\beta \\ \pm\frac{1}{2}\beta & 1 \end{pmatrix}
\tag{3}
$$

Here the off-diagonal coefficient $\beta = \Delta\rho/\rho$ is the anisotropy magnetoresistance constant that causes non-uniform current flow in the device channel. For (Ga,Mn)As epilayers we measured, its value is ~0.02. In a more general situation, equation (1) must be used to compute the resistivity tensor.

To solve for the electrical potential and current density distribution along the device channel, the following differential equations should apply,

$$
\nabla \cdot \mathbf{j} = 0
\tag{4}
$$

$$
\nabla \times \mathbf{E} = 0.
\tag{5}
$$

Ohm's law couples the two-component vectors – electrical field and current density,

$$\mathbf{E} = \hat{\rho}\mathbf{j} \tag{6}$$

Supplemental boundary conditions are,

$$j_y(x = \pm\infty, y) = 0$$
$$j_x(x = \pm\infty, y) = j \tag{7}$$
$$j_y(x, y = \pm w/2) = 0$$

In electromagnetism, a scalar potential field is usually employed to describe the system. The boundary conditions in Eq. 7 favors definition of a stream function, $\psi(x, y)$, based upon conservation of current flow,

$$j_x = \frac{\partial \psi}{\partial y}, j_y = -\frac{\partial \psi}{\partial x} \tag{8}$$

The simplest resistivity tensor (Eq. 2) yields the following equation,

$$\left(1 - \frac{\beta}{2}\cos 2\varphi\right)\frac{\partial^2 \psi}{\partial x^2} + \left(1 + \frac{\beta}{2}\cos 2\varphi\right)\frac{\partial^2 \psi}{\partial y^2} - \sin 2\varphi \frac{\partial^2 \psi}{\partial x \partial y} = 0 \tag{9}$$

By replacing the variables,

$$\xi = x - x_0 - y/k$$
$$\eta = y \tag{10}$$

The expression of $\varphi$ can be greatly simplified as

$$\varphi(\xi) = \phi_0 \operatorname{sgn}(\xi) + \delta \tag{11}$$

and Eq (9) can be rewritten in the form

$$\left(1 + k^{-2} + \sin 2\varphi k^{-1} - \frac{\beta}{2}\cos 2\varphi\right)\frac{\partial^2 \psi}{\partial \xi^2} + \left(1 + \frac{\beta}{2}\cos 2\varphi\right)\frac{\partial^2 \psi}{\partial \eta^2}$$
$$- \left[2k^{-1}\left(1 + \frac{\beta}{2}\cos 2\varphi\right) + \sin 2\varphi\right]\frac{\partial^2 \psi}{\partial \xi \partial \eta} = 0 \tag{12}$$

The boundary conditions in the form of $\psi$ are

$$\left.\frac{\partial \psi}{\partial \xi}\right|_{x=\pm\infty} = 0, \left.\frac{\partial \psi}{\partial \eta}\right|_{x=\pm\infty} = j, \left.\frac{\partial \psi}{\partial \xi}\right|_{y=\pm w/2} = 0. \tag{13}$$

We next write the flow field in two terms:

$$\psi = \psi^{(0)} + \psi^{(1)}, \ \psi^{(0)} = j\eta \tag{14}$$

The first term represents the uniform current flow without the presence of the domain wall. The second term $\psi^{(1)}$, corresponding to the perturbation due to the domain wall, should also satisfy equation (9) but with simplified boundary conditions:

$$\left.\frac{\partial \psi^{(1)}}{\partial \xi}\right|_{x=\pm\infty} = 0, \left.\frac{\partial \psi^{(1)}}{\partial \eta}\right|_{x=\pm\infty} = 0, \left.\frac{\partial \psi^{(1)}}{\partial \xi}\right|_{y=\pm w/2} = 0. \tag{15}$$

We can further separate the variables and write $\psi^{(1)}$ in the form:

$$\psi^{(1)} = \sum_{n=0}^{\infty} f_n(\xi) \cos \pi (2n+1)\eta \tag{16}$$

Here $f_n(x)$ should satisfy:

$$\left(1 + k^{-2} + \sin 2\varphi k^{-1} - \frac{\beta}{2}\cos 2\varphi\right) f^{''}(\xi) - \alpha_n^2 \left(1 + \frac{\beta}{2}\cos 2\varphi\right) f(\xi) = 0 \tag{17}$$

We seek solutions of $f_n(x)$ for $\xi > 0$ and $\xi < 0$ region individually:

$$f_n(\xi) = f_n(0) \exp\left[-\alpha_n \left(\frac{1 + \frac{\beta}{2}\cos 2\varphi}{1 + k^{-2} + \sin 2\varphi k^{-1} - \frac{\beta}{2}\cos 2\varphi}\right)^{1/2} |\xi|\right] \tag{18}$$

with,

$$f_n(0) = \frac{4 a_n^{-2} \beta j w (-1)^{n+1}}{\left(\dfrac{1 + \frac{\beta}{2}\cos 2\varphi_1}{1 + k^{-2} + \sin 2\varphi_1 k^{-1} - \frac{\beta}{2}\cos 2\varphi_1}\right)^{-1/2} + \left(\dfrac{1 + \frac{\beta}{2}\cos 2\varphi_2}{1 + k^{-2} + \sin 2\varphi_2 k^{-1} - \frac{\beta}{2}\cos 2\varphi_2}\right)^{-1/2}} \tag{19}$$

The final solution of the stream function then can be calculated,

$$\psi(x,y) = jy +$$

$$\frac{4\beta jw}{\left(\frac{1+\frac{\beta}{2}\cos 2\varphi_1}{1+k^{-2}+\sin 2\varphi_1 k^{-1}-\frac{\beta}{2}\cos 2\varphi_1}\right)^{-1/2}+\left(\frac{1+\frac{\beta}{2}\cos 2\varphi_2}{1+k^{-2}+\sin 2\varphi_2 k^{-1}-\frac{\beta}{2}\cos 2\varphi_2}\right)^{-1/2}}\sum_{n=0}^{\infty}\frac{(-1)^{n+1}\exp\left[-\alpha_n\left(\frac{1+\frac{\beta}{2}\cos 2\varphi}{1+k^{-2}+\sin 2\varphi k^{-1}-\frac{\beta}{2}\cos 2\varphi}\right)^{1/2}|x-x_0-y/k|\right]}{\alpha_n^2}\cos\left(\frac{\pi(2n+1)y}{w}\right)$$

$$(20)$$

In the simplest case where a vertical wall is considered ($k^{-1} = 0$) in a strictly cubic material with perfect alignment of device channel with cubic hard axes, $\varphi_1 = -\varphi_2 = 45^o$, the stream function has a less complex expression,

$$\psi(x,y) = jy + \frac{2\beta jw}{\pi^2}\sum_{n=0}^{\infty}\frac{(-1)^{n+1}}{(2n+1)^2}\exp\left(-\frac{\pi(2n+1)|x|}{w}\right)\cos\left(\frac{\pi(2n+1)y}{w}\right) \qquad (21)$$

## II. Current Distribution Near a Domain Wall

The current density can be easily calculated from the stream function $\psi(x,y)$:

$$j = (j_x, j_y) = \left(\frac{\partial\psi}{\partial y}, -\frac{\partial\psi}{\partial x}\right) \qquad (22)$$

This current density can be decomposed into two parts: a constant zero-order current density $j$ (Fig. 2b) and a static eddy-like current (Fig. 2c):

$$j = (j,0) + \left(\frac{\partial\psi^{(1)}}{\partial y}, -\frac{\partial\psi^{(1)}}{\partial x}\right) \qquad (23)$$

The second term, arising from the perturbation of the domain wall, is on the order of $\beta$, and is therefore usually two orders of magnitude smaller than the uniform flow part. Note that this "eddy-like" current arises from perturbations to the current streamlines in the vicinity of the resistivity discontinuity that occurs at the domain wall. It persists even for the case of a *static* domain wall and, hence, is distinct from true eddy currents that arise from domain wall motion.[13]

The calculated eddy-like distributions for vertical and tilted walls are shown in Fig. 2d and 2e, respectively. The central axis of the eddy-like current distribution is precisely centered upon the domain wall and, therefore, moves in synchrony with it. When the domain wall passes the probes, this eddy-like current distribution introduces a significant perturbation to both the transverse and longitudinal resistances, as we shall demonstrate in the next sections. This effect is clearly manifested in experimental data.[19]

### III. Transverse Resistance Generated by a Domain Wall

Our next tasks are to calculate the transverse and longitudinal resistance as a function of the domain wall position. In this section we focus on the former. The transverse voltage as a function of $x$ is,

$$V_H(x) = -\int_{-w/2}^{w/2} E_y(x, y) dy \tag{24}$$

For a vertical wall, the stream function is described by Eq. (21), the analytical expression for the transverse resistance can be written in a rather simple form,

$$R_H(x) = -\mathrm{sgn}(x - x_0)\Delta\rho\left[1 - 8\Gamma(x - x_0)\right]$$
$$= -\mathrm{sgn}(x - x_0)\frac{\Delta\rho}{2}\left\{1 - \frac{4}{\pi^2}\left[\mathrm{dilog}\left(1 - e^{-\frac{\pi|x-x_0|}{w}}\right) - \mathrm{dilog}\left(1 + e^{-\pi\frac{\pi|x-x_0|}{w}}\right)\right]\right\} \tag{25}$$

Here we employ

$$\sum_{n=0}^{\infty} \frac{1}{\pi^2 (2n+1)^2} \exp\left(-\pi(2n+1)|x|\right) = \frac{1}{2\pi^2}\left[di\log\left(1-e^{-\pi|x|}\right) - di\log\left(1+e^{-\pi|x|}\right)\right] = \Gamma(x)$$

(26)

The transverse resistance calculated for a vertical domain wall is plotted in Fig. 3a. It is evident that the perturbation from such an abrupt change is significant. However, even for this case the domain wall signal spans a length comparable to a significant fraction of width of the device channel. It decays exponentially beyond that length scale. It is worth mentioning that even though the magnitude of the signal depends on the anisotropy ratio $\beta$, the spatial distribution of the transverse voltage is independent of this parameter. It is determined by the broken symmetry arising from the presence of the domain wall.

Figure 3a also shows the transverse voltage profiles for domain walls tilted at $30^o$ and $60^o$. Comparing these results to those generated from a vertical wall makes it clear that only the regions that the domain wall physically spans are affected. Changes from a vertical wall to one with finite slope result in a linear extension of the signal. Hence, depending on the slope of the wall, signals varying from a sigmoidal to linear shape are expected as the domain wall traverses transverse probes. Domain walls of various slopes have been observed in experiments; several examples of such traces are shown in Fig. 3b.[20] These data are recorded when a domain wall is driven at a constant velocity with a fixed external magnetic field. The horizontal axis is determined by scaling with the measured domain wall velocity.[20] The domain wall shape in the device channel demonstrates a striking dependence on the external field orientation. When the external field is orientated along the device channel (the magnetic hard axis), a rather extended domain wall signature (reflecting small slope) is usually nucleated and swept across the sample. On the other hand, when the external field is tilted away from the longitudinal axis of the device, a domain wall with sigmoidal form is manifested in the transverse signals; this reflects a DW orientation that is closer to vertical (large slope).

When a domain wall moves close enough to a transverse probe pair – specifically, within a distance comparable to the width of the device channel – a giant planar Hall resistance can be detected. Once the domain wall slope is known, this measured GPHE signal can be used to determine the domain wall position. In practice, calculations shows that the GPHE signal does not change significantly when the slope corresponds to an angle larger than about 70 degrees. If the external field is applied closer to the easy as opposed to the hard axes, the domain wall slope falls into this angular regime. Hence, for such cases, Eq. 25 is adequate for predicting the domain wall position.

## IV. Longitudinal Resistance: Protocol for Deducing the "Intrinsic" Domain Wall Resistance

Interpretation of the longitudinal resistance becomes rather complicated when a domain wall is present in the device channel. In experiments it is found that the longitudinal resistance measured from the top probe pair ($R_{xx}^{U} = R_{14,23}$) is different than that obtained from the bottom pair ($R_{xx}^{D} = R_{14,65}$). The model makes clear that this difference arises from the complex current distribution around the domain wall, as depicted in Figures 2d and 2e. For each fixed position of the domain wall, $x_0$, the potential sum rule relates the difference in longitudinal resistances to the difference in across the transverse probes,

$$\delta R(x_0) \ = \ R_{xx}^{U}(x_0) - R_{xx}^{D}(x_0) \ = \ R_{xy}^{L}(x_0) - R_{xy}^{R}(x_0). \tag{27}$$

Here $R_{xy}^L \equiv R_{14,26}$ and $R_{xy}^R \equiv R_{14,35}$ are the transverse resistances measured at the left and right probes, as presented earlier. As seen, the transverse resistance adds a contribution that has opposite sign for the top and bottom longitudinal resistances, and their respective values depend upon the position of the domain wall within the device.

The voltage developed along the top and bottom of the device channel can be calculated from the stream function obtained above,

$$V^{U,D} = -\int\limits_{-a}^{a} E_x(x, y = \pm 1/2) dx = -\int\limits_{-a}^{a} \frac{\partial \psi}{\partial y} dx .\qquad(28)$$

In general, analytical expression for the longitudinal resistances can be derived from the stream function obtained in Eq. 20. Again, we first consider the simplest case where a vertical domain wall is measured in a sample with perfect alignment of device channel along the magnetic hard axes,

$$R^{U,D} = \bar{\rho}\frac{2a}{w} \pm 4\Delta\rho \Big[ \text{sgn}\left(x_0 + a\right)\Gamma\left(x_0 + a\right) - \text{sgn}\left(x_0 - a\right)\Gamma\left(x_0 - a\right) \Big].\qquad(29)$$

The second term describes the perturbation that results from an admixture of transverse resistance resulting from the presence of the domain wall. Apparently, this expression and the transverse resistance (Eq. 25) both satisfy the sum rule of Eq. 27. It is evident from this equation that this contribution from the transverse resistance can be cancelled by averaging $R^U$ and $R^D$, $\bar{R} = \left(R^U + R^D\right)\big/2$ since it contributes with different sign for $R^U$ and $R^D$.

Figure 4a displays the zero-field longitudinal resistances, $R_{xx}^U$ and $R_{xx}^D$, and their average $\bar{R}_{xx}$, versus the position of a vertical DW as it is stepped along a well-aligned device channel. For these plots the aspect ratio (length:width) is assumed to be 6:1. Values on the *y*-axis represent resistance deviation from values in the absence of a domain wall. The red and blue curves represent measurement from the top and bottom pairs of electrodes, respectively, while the black curve is their average. The initial decrease (increase) in $R_{xx}^U$ ($R_{xx}^D$) is half of the transverse resistance, as is evident from Eq. 29. Note that $x_0 \sim -a$, $\Gamma(x_0 - a) \to 0$, and

$$\delta R^{U,D} = \pm 4\Delta\rho\, \mathrm{sgn}(x_0 + a)\,\Gamma(x_0 + a) = \mp R_H(x_0 + a)/2\,. \qquad (30)$$

When the domain wall is fully contained between the probes, the longitudinal resistances maintain constant values.

For domain walls with arbitrary tilt angle, the equations are integrated numerically. Results for a DW of 75° tilt angle are plotted in Fig. 4b. Except for the bump and dip that appear when domain wall passes the probes, the transverse admixture to the longitudinal resistances are seen to be completely compensated (*i.e.* nulled) by averaging the measurements from top and bottom of the device channel.

The results from these calculations are consistent with what is observed in experiment.[19] The experimentally observed values for $R_{xx}^U$ and $R_{xx}^D$ follow the theoretical predictions quite closely. The presence of the small bump and dip in $\bar{R}_{xx}$ reflect the presence of a domain wall that is tilted with respect to the channel. The magnitudes of these features can be employed to deduce directly the tilt angle of the domain wall (represented in Fig. 1). The calculated average resistances ($\overline{R}_{xx}$) from domain walls with various tilt angles are plotted in Fig. 4c. The size of the bump and dip becomes smaller as the domain wall angle approaches 90$^\text{o}$. These features vanish when the domain is precisely vertical. The correspondence between the magnitude of the bump and dip and the domain wall slope is presented in the inset of Fig. 4c. The slope of the domain wall present in experiments can be determined from this curve.

This scenario changes slightly when there is slight misalignment of magnetic hard axis with respect to the orientation of the device channel. Due to the non-uniformity of the materials and the slight imprecision of the fabrication process, such small misalignments are, in general, inevitable. Fig. 4d shows the average resistance calculated from a device channel oriented with 0.03$^\text{o}$ misalignment with respect to the hard axes. The asymptotic value is extremely small, on the order of 10$^{-5}$ of the sheet resistance. This tiny asymptotic value can be compensated by linearly interpolating between the asymptotic values of the longitudinal resistance. Results for domain wall slope of 80$^\text{o}$, 85$^\text{o}$ and 90$^\text{o}$ are shown for comparison. The vertical wall (90$^\text{o}$) matches this simple interpolation scheme. When the domain wall is not vertical, the simple linear interpolation remains a good description of the resistances for the centermost region of the device (between the transverse probes). When the domain wall is in the vicinity of these probes – that is when the domain wall is entering or leaving the measurement region – the bump and dip are manifested. In this regime deviations from the linear interpolation are seen to occur.

### V. Conclusions

We have developed a simple model for the longitudinal and transverse resistances arising within multiterminal (Ga,Mn)As device containing a single domain wall. This model accounts for the classical electrical potential distribution expected in the vicinity of domain wall as a consequence of the local discontinuity in resistivity. It elucidates the separate contributions to the longitudinal resistances that originate from the presence of this discontinuity. An important clarification emerges from this analysis: averaging the longitudinal resistances measured from the top and bottom of the device channel ($R_{xx}^{U}$ and $R_{xx}^{D}$, respectively) nulls the inadvertent admixture of transverse resistance arising from the presence of a domain wall within the device. Without such clarification, this might ostensibly appear to be an effect reflecting from local magnetic scattering at the domain wall, i.e. the "intrinsic" domain wall resistance. It is not.

An intrinsic domain wall resistance, if present, will contribute identically to $R_{xx}^{U}$ and $R_{xx}^{D}$, and should not depend on the position of the domain wall. From the analysis herein it becomes clear how the intrinsic domain wall resistance, if present, can be deduced from the experimental data.. The average resistance that is measured experimentally, $\bar{R}_{xx}^{(\exp)} = ( \ [\bar{R}_{xx}^{U}]^{(\exp)} + [\bar{R}_{xx}^{U}]^{(\exp)} \ ) / 2$, can be compared with the average resistance predicted by the model presented here, $\bar{R}_{xx}$. The model accounts *solely* for potentials associated with the resistivity discontinuity at the domain wall. In regions between (but not too close) to the transverse probe pairs the model predicts a linearly-evolving average longitudinal resistance (which is everywhere identically zero for a precisely-aligned device channel). In this region, differences between experimental data and the model's predictions, $R_{DW} = \overline{R}_{xx}^{(\text{exp})} - \overline{R}_{xx}$, should directly reflect the role of local magnetic scattering at the domain wall.

Elsewhere[19] we have employed the analysis described here to deduce the domain wall resistance in a family of carefully-patterned multiterminal (Ga,Mn)As devices. For the epilayers employed in these experiments we find that a consistently negative intrinsic value is obtained. This may reflect quantum corrections to the resistivity due to magnetic scattering at the domain wall.

## Acknowledgements

We acknowledge support from DARPA under grants DSO/SPINS-MDA 972-01-1-0024. We also thank Leon Balents, Anton Burkov, and Martin Veillette for initial discussions on this topic.

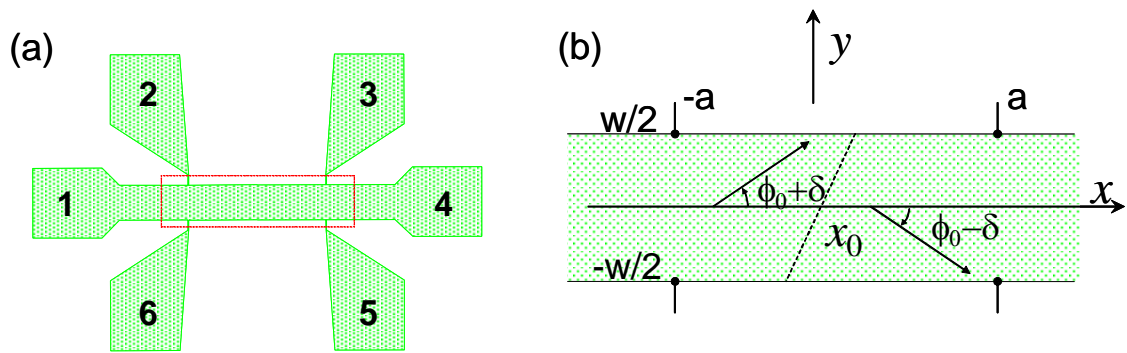

**Fig. 1.** Model of the experimental geometry. (**a**) Plan view of entire device showing current probes (1,4) and voltage probes (2,3,5,6). (**b**) The domain wall is situated in the device channel (shaded green) at position $x_0$ with slope $k$ with respect to the channel walls. Four voltage probes located at ($\pm a$, $\pm w/2$) are used to measured the transverse and longitudinal resistances defined in the text.

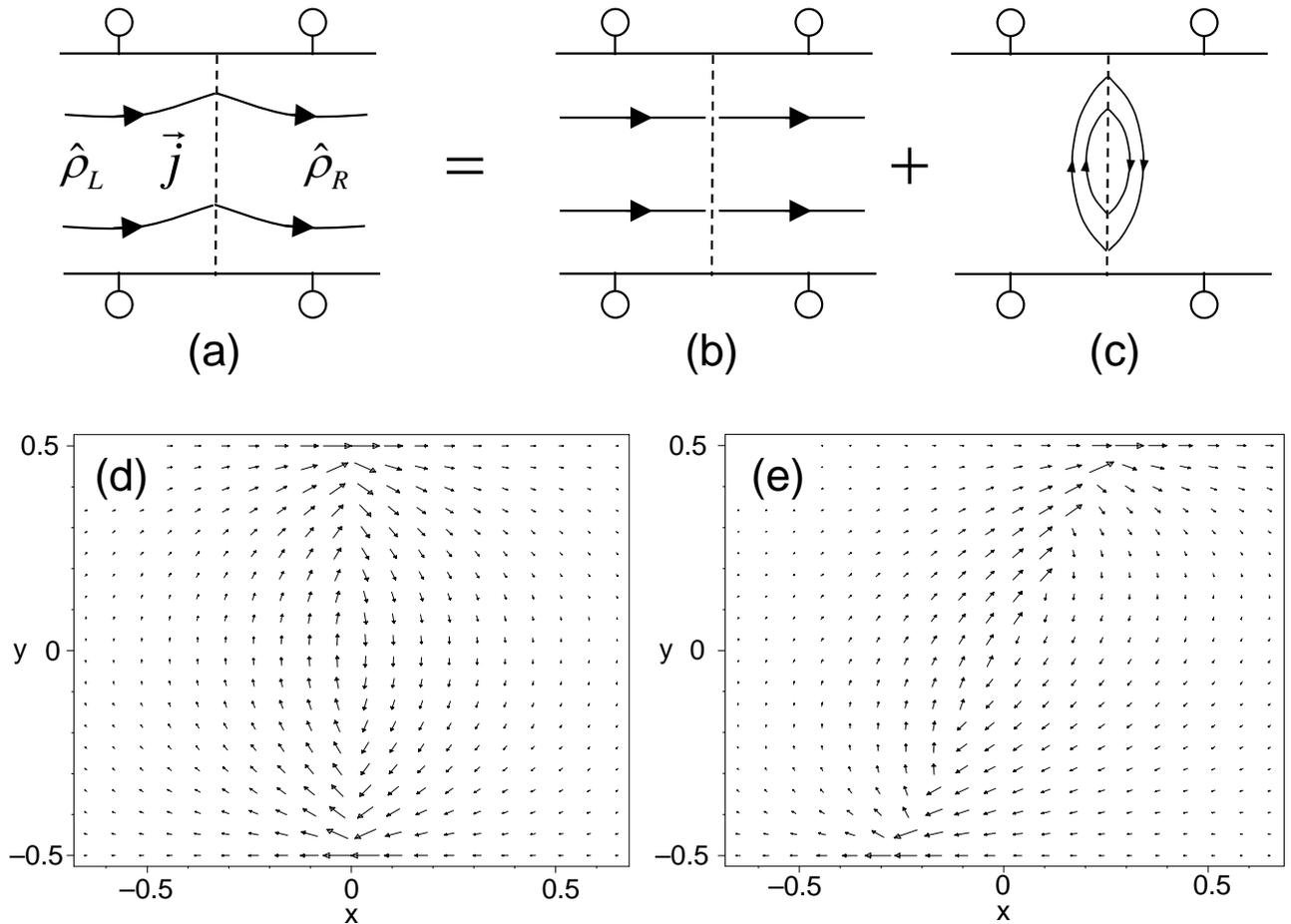

(a)          (b)          (c)

(d)          (e)

**Fig. 2**. Static eddy-like current distribution near a domain wall. **(a,b,c)** The perturbation of the stationary magnetic domain wall causes a static non-uniform current distribution within the device channel. It can be decomposed into a uniform current flow as in **(b)** and a vortex-like current distribution around the domain wall as plotted **(c)**. **(d)** The eddy-like current distribution around a vertical magnetic domain wall. **(e)** The eddy-like current distribution around a domain wall with a slope of $60^{\circ}$. For clarity, the magnitude of vectors shown in figures c, d, and e are magnified by two orders of magnitude.

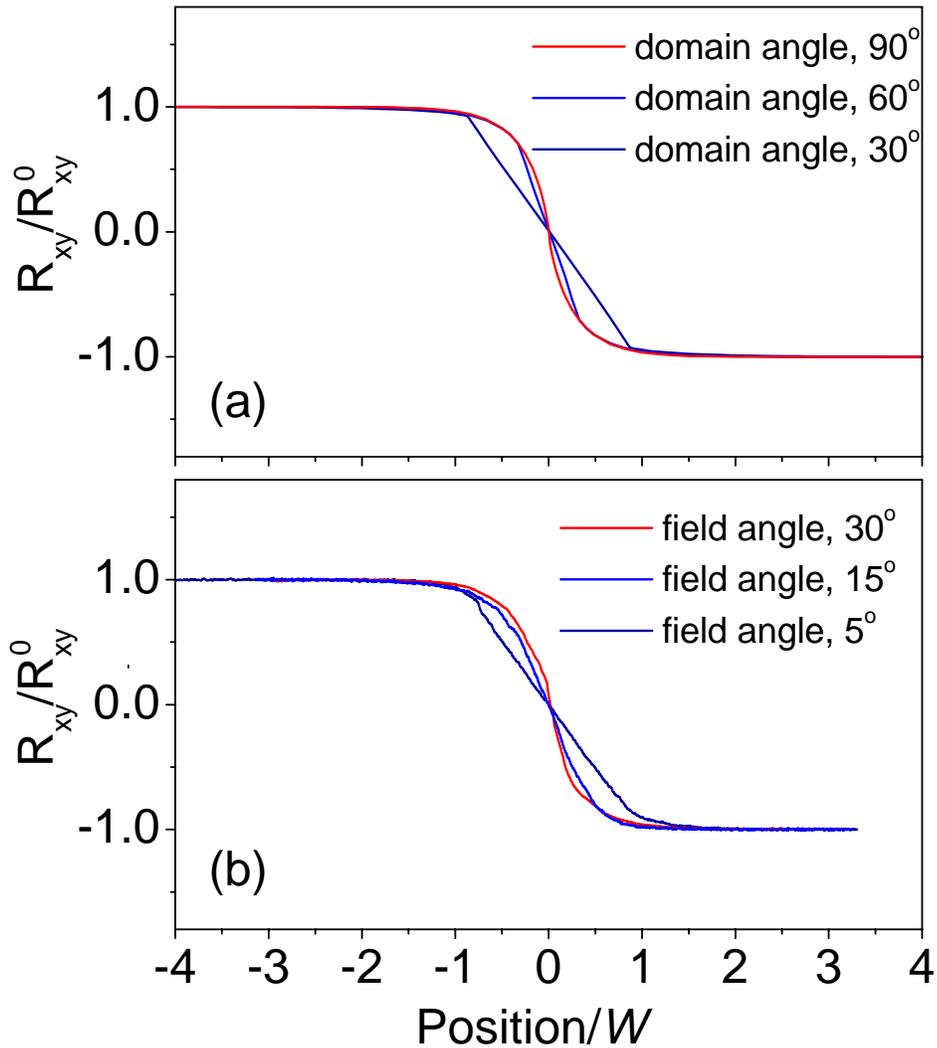

**Fig. 3**. **(a)** Calculated transverse voltage profile for domain walls possessing different slopes with respect the longitudinal device axis. The transverse resistances, $R_{xy}$, are normalized to its saturated value; $R_{xy}^0$; the position of the domain wall is normalized to the domain wall width, *W*. **(b)** Experimentally observed transverse signal generated by a single domain wall as it passes by a transverse probe pair. The data are obtained with different driving field angles from [110] orientation [After Ref. 20].

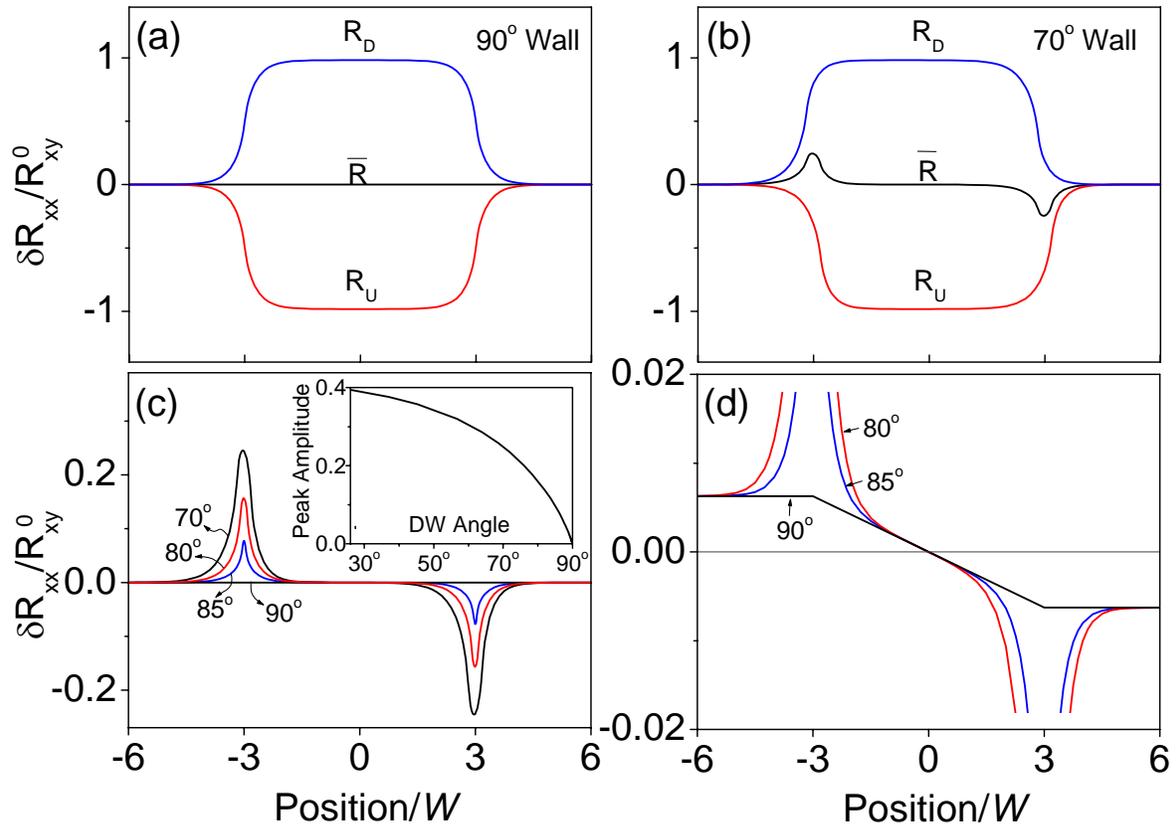

**Fig. 4. (a,b)** Calculated longitudinal resistances at the top and bottom of a device channel, and their average as a function of the position of domain wall when a perfect alignment is achieved between the longitudinal device axis and crystallographic [110] direction. Results shown are for a vertical domain wall (a) and a tilted domain wall (b), and graph axes are normalized as in Fig. 3. Except for the bump and dip that appear when domain wall passes the probes, the planar Hall admixture to the longitudinal measurements are completely compensated by averaging the measurements from top and bottom of the device channel. **(c)** A magnified view of the average resistance, as a function of domain wall angle of slope. **(d)** A small misalignment of 0.03 degree will yield a tiny asymptotic value in the average resistance. The interpolation between the asymptotic values gives a good description of the resistances from the domain effect.